\documentclass[12pt]{article}
  \usepackage{amsfonts}
  \usepackage{amsmath}
\usepackage{amssymb}
\usepackage{amscd}
\usepackage[dvips]{graphicx}
\begin{document}

~~
\bigskip
\bigskip
\begin{center}
{\Large {\bf{{{Landau energy spectrum and quantum oscillator model for twisted N-enlarged Newton-Hooke space-time}}}}}
\end{center}
\bigskip
\bigskip
\bigskip
\begin{center}
{{\large ${\rm {Marcin\;Daszkiewicz}}$}}
\end{center}
\bigskip
\begin{center}
\bigskip

{ ${\rm{Institute\; of\; Theoretical\; Physics}}$}

{ ${\rm{ University\; of\; Wroclaw\; pl.\; Maxa\; Borna\; 9,\;
50-206\; Wroclaw,\; Poland}}$}

{ ${\rm{ e-mail:\; marcin@ift.uni.wroc.pl}}$}

\end{center}
\bigskip
\bigskip
\bigskip
\bigskip
\bigskip
\bigskip
\bigskip
\bigskip
\bigskip
\begin{abstract}
We derive the energy levels for oscillator model defined on the twisted N-enlarged Newton-Hooke space-time, i.e., we find  time-dependent
eigenvalues and  corresponding time-dependent eigenstates. We also demonstrate that for a particular choice of deformation parameters of phase space the above spectrum can be identified with the time-dependent Landau one.
\end{abstract}
\bigskip
\bigskip
\bigskip
\bigskip
\eject

\section{{{Introduction}}}

The suggestion to use noncommutative coordinates goes back to
Heisenberg and was firstly  formalized by Snyder in \cite{snyder}.
Recently, there were also found formal  arguments based mainly  on
Quantum Gravity \cite{2}, \cite{2a} and String Theory models
\cite{recent}, \cite{string1}, indicating that space-time at Planck
scale  should be noncommutative, i.e., it should  have a quantum
nature. Consequently, there are a number of papers dealing with
noncommutative classical and quantum  mechanics (see e.g.
\cite{mech}-\cite{qm}) as well as with field theoretical models
(see e.g. \cite{prefield}-\cite{fiorewess}) in which  the quantum
space-time is employed.

It is well-known that  a proper modification of the Poincare and
Galilei Hopf algebras can be realized in the framework of Quantum
Groups \cite{qg1}, \cite{qg3}. Hence, in accordance with the
Hopf-algebraic classification  of all deformations of relativistic
and nonrelativistic symmetries (see \cite{class1}, \cite{class2}),
one can distinguish three
types of quantum spaces \cite{class1}, \cite{class2} (for details see also \cite{nnh}):\\
\\
{ \bf 1)} Canonical ($\theta^{\mu\nu}$-deformed) type of quantum space \cite{oeckl}-\cite{dasz1}
\begin{equation}
[\;{ x}_{\mu},{ x}_{\nu}\;] = i\theta_{\mu\nu}\;, \label{noncomm}
\end{equation}
\\
{ \bf 2)} Lie-algebraic modification of classical space-time \cite{dasz1}-\cite{lie1}
\begin{equation}
[\;{ x}_{\mu},{ x}_{\nu}\;] = i\theta_{\mu\nu}^{\rho}{ x}_{\rho}\;,
\label{noncomm1}
\end{equation}
and\\
\\
{ \bf 3)} Quadratic deformation of Minkowski and Galilei  spaces \cite{dasz1}, \cite{lie1}-\cite{paolo}
\begin{equation}
[\;{ x}_{\mu},{ x}_{\nu}\;] = i\theta_{\mu\nu}^{\rho\tau}{
x}_{\rho}{ x}_{\tau}\;, \label{noncomm2}
\end{equation}
with coefficients $\theta_{\mu\nu}$, $\theta_{\mu\nu}^{\rho}$ and  $\theta_{\mu\nu}^{\rho\tau}$ being constants.\\
\\
Moreover, it has been demonstrated in \cite{nnh}, that in the case of the 
so-called N-enlarged Newton-Hooke Hopf algebras
$\,{\mathcal U}^{(N)}_0({ NH}_{\pm})$ the twist deformation
provides the new  space-time noncommutativity of the
form\footnote{$x_0 = ct$.},\footnote{ The discussed space-times have been  defined as the quantum
representation spaces, so-called Hopf modules (see e.g. \cite{oeckl}, \cite{chi}), for the quantum N-enlarged
Newton-Hooke Hopf algebras.}
\begin{equation}
{ \bf 4)}\;\;\;\;\;\;\;\;\;[\;t,{ x}_{i}\;] = 0\;\;\;,\;\;\; [\;{ x}_{i},{ x}_{j}\;] = 
if_{\pm}\left(\frac{t}{\tau}\right)\theta_{ij}(x)
\;, \label{nhspace}
\end{equation}
with time-dependent  functions
$$f_+\left(\frac{t}{\tau}\right) =
f\left(\sinh\left(\frac{t}{\tau}\right),\cosh\left(\frac{t}{\tau}\right)\right)\;\;\;,\;\;\;
f_-\left(\frac{t}{\tau}\right) =
f\left(\sin\left(\frac{t}{\tau}\right),\cos\left(\frac{t}{\tau}\right)\right)\;,$$
$\theta_{ij}(x) \sim \theta_{ij} = {\rm const}$ or
$\theta_{ij}(x) \sim \theta_{ij}^{k}x_k$ and  $\tau$ denoting the time scale parameter
 -  the cosmological constant. Besides, it should be  noted that the mentioned above quantum spaces {\bf 1)}, { \bf 2)} and { \bf 3)}
 can be obtained  by the proper contraction limit  of the commutation relations { \bf 4)}\footnote{Such a result indicates that the twisted N-enlarged Newton-Hooke Hopf algebra plays a role of the most general type of quantum group deformation at nonrelativistic level.}.

 As mentioned above, recently, there has been discussed the impact of different  kinds of
quantum spaces on the dynamical structure of physical systems (see e.g. \cite{mech}-\cite{field} and \cite{romero}-\cite{giri}).
Particulary, it has been demonstrated, that
in the case of a classical oscillator model \cite{kijanka} as well as in the case of a nonrelativistic particle moving in constant
external field force $\vec{F}$ \cite{daszwal}, there are generated by space-time noncommutativity additional force terms. Such a type
of investigation has been  performed for quantum oscillator model as well \cite{kijanka}, i.e., it was demonstrated that the  quantum space in nontrivial way affects  the spectrum of the energy operator. Besides, in the paper \cite{toporzelek} there has been considered a model of a particle moving on the $\kappa$-Galilei space-time in the presence of gravitational field force. It has been demonstrated, that in such a case
there is produced a force term, which can be identified with the so-called Pioneer anomaly \cite{pioneer}, and the value of the deformation parameter $\kappa$ can be fixed by a comparison of obtained result with observational data.
Moreover,  especially interesting results have been obtained in the series of papers
\cite{hallcan}-\cite{hallcanf} concerning the Hall effect for canonically deformed space-time (\ref{noncomm}). Particularly, there has been found the $\theta$-dependent (Landau)
energy spectrum of an electron moving in uniform magnetic as well as in uniform electric field. Such results have been generalized to the case
of the twisted N-enlarged Newton-Hooke Hopf algebra \cite{nnh}, i.e., there has been derived the time-dependent Landau levels for the particle moving in the
corresponding quantum space in the presence of external fields \cite{daszlandau}.


In this article we find the time-dependent energy levels for oscillator model defined on noncommutative space-time (\ref{spaces}).
In particular, we  demonstrate that for a special choice of deformation parameters of corresponding phase space (see formula (\ref{condition}))
the above spectrum can be identified with time-dependent Landau one.

The paper is organized as follows. In Sect. 2 we recall basic facts
concerning the twisted N-enlarged Newton-Hooke   space-times
provided in the article \cite{nnh}. The third section is devoted to the calculation of energy spectrum (as well as Landau levels) for the twist-deformed oscillator model. The
final remarks are presented in the last section.

\section{Twisted N-enlarged Newton-Hooke space-times}

In this section we recall the basic facts associated with the twisted N-enlarged Newton-Hooke Hopf algebra $\;{\cal U}^{(N)}_{\alpha}(NH_{\pm})$ and with the
corresponding quantum space-times \cite{nnh}.  Firstly, it should be noted, that in accordance with Drinfeld  twist procedure, the algebraic sector of twisted
Hopf structure $\;{\cal U}^{(N)}_{\alpha}(NH_{\pm})$ remains
undeformed, i.e., it takes the form
 \begin{eqnarray}
&&\left[\, M_{ij},M_{kl}\,\right] =i\left( \delta
_{il}\,M_{jk}-\delta _{jl}\,M_{ik}+\delta _{jk}M_{il}-\delta
_{ik}M_{jl}\right)\;\; \;, \;\;\; \left[\, H,M_{ij}\,\right] =0
 \;,  \label{q1} \\
&~~&  \cr &&\left[\, M_{ij},G_k^{(n)}\,\right] =i\left( \delta
_{jk}\,G_i^{(n)}-\delta _{ik}\,G_j^{(n)}\right)\;\; \;, \;\;\;\left[
\,G_i^{(n)},G_j^{(m)}\,\right] =0 \;,\label{q2}
\\
&~~&  \cr &&\left[ \,G_i^{(k)},H\,\right] =-ikG_i^{(k-1)}\;\; \;, \;\;\; \left[\, H,G_i^{(0)}\,\right] =\pm \frac{i}{\tau}G_i^{(1)}\;\;\;;\;\;\;k>1
\;,\label{q3}
\end{eqnarray}
where $\tau$, $M_{ij}$, $H$, $G_i^{(0)} (=P_i)$, $G_i^{(1)} (=K_i)$ and $G_i^{(n)} (n>1)$ can be identified with cosmological time parameter, rotation, time translation, momentum, boost and accelerations  operators respectively. Besides, the coproducts and antipodes of considered algebra are given by\footnote{$\Delta_0(a) = a\otimes 1 +1\otimes a$, $S_{0}(a) =-a$.}
\begin{equation}
 \Delta _{\alpha }(a) = \mathcal{F}_{\alpha }\circ
\,\Delta _{0}(a)\,\circ \mathcal{F}_{\alpha }^{-1}\;\;\;,\;\;\;
S_{\alpha}(a) =u_{\alpha }\,S_{0}(a)\,u^{-1}_{\alpha }\;,\label{fs}
\end{equation}
with $u_{\alpha }=\sum f_{(1)}S_0(f_{(2)})$ (we use the Sweedler's notation
$\mathcal{F}_{\alpha }=\sum f_{(1)}\otimes f_{(2)}$) and with the twist factor
$\mathcal{F}_{\alpha } \in {\cal U}^{(N)}_{\alpha}(NH_{\pm}) \otimes
{\cal U}^{(N)}_{\alpha}(NH_{\pm})$
satisfying  the classical cocycle condition
\begin{equation}
{\mathcal F}_{{\alpha }12} \cdot(\Delta_{0} \otimes 1) ~{\cal
F}_{\alpha } = {\mathcal F}_{{\alpha }23} \cdot(1\otimes \Delta_{0})
~{\mathcal F}_{{\alpha }}\;, \label{cocyclef}
\end{equation}
and the normalization condition
\begin{equation}
(\epsilon \otimes 1)~{\cal F}_{{\alpha }} = (1 \otimes
\epsilon)~{\cal F}_{{\alpha }} = 1\;, \label{normalizationhh}
\end{equation}
such that ${\cal F}_{{\alpha }12} = {\cal F}_{{\alpha }}\otimes 1$ and
${\cal F}_{{\alpha }23} = 1 \otimes {\cal F}_{{\alpha }}$.

The corresponding quantum space-times are defined as the representation spaces (Hopf modules) for the N-enlarged Newton-Hooke Hopf algebra
\;${\cal U}_{\alpha}^{(N)}(NH_{\pm})$. Generally, they are equipped with two the spatial directions
commuting to classical time, i.e. they take  the form
\begin{equation}
[\;t,\hat{x}_{i}\;] =[\;\hat{x}_{1},\hat{x}_{3}\;] = [\;\hat{x}_{2},\hat{x}_{3}\;] =
0\;\;\;,\;\;\; [\;\hat{x}_{1},\hat{x}_{2}\;] =
if({t})\;\;;\;\;i=1,2,3
\;. \label{spaces}
\end{equation}
 However, it should be noted
that this type of noncommutativity  has  been  constructed explicitly  only in the case of the 6-enlarged Newton-Hooke Hopf algebra, with
\cite{nnh}\footnote{$\kappa_a = \alpha$ $(a=1,...,36)$ denote the deformation parameters.}
\begin{eqnarray}
f({t})&=&f_{\kappa_1}({t}) =
f_{\pm,\kappa_1}\left(\frac{t}{\tau}\right) = \kappa_1\,C_{\pm}^2
\left(\frac{t}{\tau}\right)\;, \nonumber\\
f({t})&=&f_{\kappa_2}({t}) =
f_{\pm,\kappa_2}\left(\frac{t}{\tau}\right) =\kappa_2\tau\, C_{\pm}
\left(\frac{t}{\tau}\right)S_{\pm} \left(\frac{t}{\tau}\right) \;,
\nonumber\\
&~~&~~~~~~~~~~~~~~~~~~~~~~~~~~~~~~~~~ \nonumber\\
&~~&~~~~~~~~~~~~~~~~~~~~~~~~~~~~~~~~~\cdot \nonumber\\
&~~&~~~~~~~~~~~~~~~~~~~~~~~~~~~~~~~~~\cdot \label{w2}\\
&~~&~~~~~~~~~~~~~~~~~~~~~~~~~~~~~~~~~\cdot \nonumber\\
&~~&~~~~~~~~~~~~~~~~~~~~~~~~~~~~~~~~~ \nonumber
\end{eqnarray}
\begin{eqnarray}
&~~&~~~~~~~~~~~~~~~~~~~~~~~~~~~~~~~~~ \nonumber\\
&~~&~~~~~~~~~~~~~~~~~~~~~~~~~~~~~~~~~\cdot \nonumber\\
&~~&~~~~~~~~~~~~~~~~~~~~~~~~~~~~~~~~~\cdot \nonumber\\
&~~&~~~~~~~~~~~~~~~~~~~~~~~~~~~~~~~~~\cdot \nonumber\\
&~~&~~~~~~~~~~~~~~~~~~~~~~~~~~~~~~~~~ \nonumber\\
f({t})&=&
f_{\kappa_{35}}\left(\frac{t}{\tau}\right) = 86400\kappa_{35}\,\tau^{11}
\left(\pm C_{\pm} \left(\frac{t}{\tau}\right)  \mp \frac{1}{24}\left(\frac{t}{\tau}\right)^4 - \frac{1}{2}
\left(\frac{t}{\tau}\right)^2 \mp 1\right) \,\times \nonumber\\
&~~&~~~~~~~~~~~~~~~~\times~\;\left(S_{\pm} \left(\frac{t}{\tau}\right)  \mp \frac{1}{6}\left(\frac{t}{\tau}\right)^3 - \frac{t}{\tau}\right)\;,
\nonumber\\
f({t})&=&
f_{\kappa_{36}}\left(\frac{t}{\tau}\right) =
518400\kappa_{36}\,\tau^{12}\left(\pm C_{\pm} \left(\frac{t}{\tau}\right)  \mp \frac{1}{24}\left(\frac{t}{\tau}\right)^4 - \frac{1}{2}
\left(\frac{t}{\tau}\right)^2 \mp 1\right)^2\;, \nonumber
\end{eqnarray}
and
$$C_{+/-} \left(\frac{t}{\tau}\right) = \cosh/\cos \left(\frac{t}{\tau}\right)\;\;\;{\rm and}\;\;\;
S_{+/-} \left(\frac{t}{\tau}\right) = \sinh/\sin
\left(\frac{t}{\tau}\right) \;.$$
Moreover, one can easily check that when $\tau$ is approaching the infinity limit the above quantum spaces reproduce the canonical (\ref{noncomm}),
Lie-algebraic (\ref{noncomm1}) and quadratic (\ref{noncomm2})  type of
space-time noncommutativity, i.e., for $\tau \to \infty$ we get
\begin{eqnarray}
f_{\kappa_1}({t}) &=& \kappa_1\;,\nonumber\\
f_{\kappa_2}({t}) &=& \kappa_2\,t\;,\nonumber\\
&\cdot& \nonumber\\
&\cdot& \label{qqw2}\\
&\cdot& \nonumber\\
f_{\kappa_{35}}({t}) &=& \kappa_{35}\,t^{11}\;, \nonumber\\
f_{\kappa_{36}}({t}) &=& \kappa_{36}\,t^{12}\;. \nonumber
\end{eqnarray}
Of course, for all deformation parameters $\kappa_a$  going to zero the above deformations disappear.

\section{Quantum oscillator model for twisted N-enlarged Newton-Hooke space-time}

Let us now return to the main aim of our investigations, i.e., to the oscillator model defined on quantum space-times (\ref{spaces})-(\ref{qqw2}).
In first step of our construction, we extend the described in pervious section spaces to the whole algebra of momentum and position operators as follows
\begin{eqnarray}
&&[\;\hat{ x}_{1},\hat{ x}_{2}\;] = 2if_{\kappa_a}({t})\;\;\;,\;\;\;
[\;\hat{ p}_{1},\hat{ p}_{2}\;] = 2ig_{\kappa_a}({t})\;,\label{rel1}\\
&&[\;\hat{ x}_{i},\hat{ x}_{j}\;] =
i\delta_{ij}\left[1+ f_{\kappa_a}({t})g_{\kappa_a}({t})\right]\;, \label{rel2}
\end{eqnarray}
with the arbitrary function $g_{\kappa_a}({t})$. One can check that relations
(\ref{rel1}), (\ref{rel2}) satisfy the Jacobi identity and for deformation parameters
$\kappa_a$ approaching zero become classical. \\
Next, by analogy to the commutative case, we define the Hamiltonian operator
\begin{eqnarray}
\hat{H} = \frac{1}{2m}\left({\hat{{p}}_1^2}+{\hat{{p}}_2^2} \right) +
\frac{1}{2}m\omega^2 \left({\hat{{x}}_1^2}+{\hat{{x}}_2^2} \right) \;. \label{2dhn}
\end{eqnarray}
with $m$ and $\omega$ denoting the mass and frequency of a particle, respectively. \\
In order to analyze the above system, we represent the 
noncommutative operators $({\hat x}_i, {\hat p}_i)$ by the classical
ones $({ x}_i, { p}_i)$ as  (see e.g.
\cite{romero1}-\cite{kijanka})
\begin{eqnarray}
{\hat x}_{1} &=& { x}_{1} - {f_{\kappa_a}(t)}p_2\;,\label{rep1}\\
{\hat x}_{2} &=& { x}_{2} +{f_{\kappa_a}(t)}p_1
\;,\label{rep2}\\
{\hat p}_{1} &=& { p}_{1} + {g_{\kappa_a}(t)}x_2\;,\label{rep2a}\\
{\hat p}_{2} &=& { p}_{2} -{g_{\kappa_a}(t)}x_1
\;,\label{rep3}
\end{eqnarray}
where
\begin{equation}
[\;x_i,x_j\;] = 0 =[\;p_i,p_j\;]\;\;\;,\;\;\; [\;x_i,p_j\;]
={i\hbar}\delta_{ij}\;. \label{classpoisson}
\end{equation}
Then, the  Hamiltonian (\ref{2dhn}) takes the form
\begin{eqnarray}
\hat{H} = \hat{H}(t) =
\frac{1}{2M(t)}\left({{{p}}_1^2}+{{{p}}_2^2} \right)  +
\frac{1}{2}M(t)\Omega^2(t)\left({{{x}}_1^2}+{{{x}}_2^2} \right)
- S(t)L\;, \label{2dh1}
\end{eqnarray}
with
\begin{eqnarray}
&&L = x_1p_2 - x_2p_1\;, \\
&&1/M(t) = 1/m +m\omega^2 f_{\kappa_a}^2(t) \;,\\
&&\Omega(t) = \sqrt{\left(1/m
+m\omega^2 f_{\kappa_a}^2(t) \right)\left(m\omega^2
+ g_{\kappa_a}^2(t) /m\right)}\;,
\end{eqnarray}
and
\begin{equation}
S(t)=m\omega^2f_{\kappa_a}(t) +g_{\kappa_a}(t)/m\;.
\end{equation}

 In accordance with the scheme proposed in  \cite{kijanka},  we introduce a set of time-dependent creation
$(a^{\dag}_{A}(t))$ and annihilation
$(a_{A}(t))$ operators 
\begin{eqnarray}
\hat{a}_{\pm}(t) &=& \frac{1}{2}\left[\frac{({p}_2 \pm
i{p}_1)}{\sqrt{M(t)\Omega (t)}} -i\sqrt{M(t)\Omega
(t)}({x}_2 \pm
i{x}_1)\right]\;,\label{oscy1}
\end{eqnarray}
satisfying the standard commutation relations
\begin{eqnarray}
[\;\hat{a}_{A},\hat{a}_{B}\;] =
0\;\;,\;\;[\;\hat{a}^{\dag}_{A},\hat{a}^{\dag}_{B}\;]
=0\;\;,\;\;[\;\hat{a}_{A},\hat{a}^{\dag}_{B}\;] =
\delta_{AB}\;\;\;;\;\;A,B = \pm\;.\label{ccr1}
\end{eqnarray}
Then, it is easy to see  that in  terms of the  objects (\ref{oscy1}) the Hamiltonian function (\ref{2dh1}) can be written as 
follows
\begin{equation}
{\hat{H}}(t)=\Omega_{+}(t) \left({\hat N}_+(t) + \frac{1}{2}\right)
+ \Omega_{-}(t) \left({\hat N}_-(t) + \frac{1}{2}\right)  \;, \label{hamquantosc}
\end{equation}
with the  coefficient  $\Omega_{\pm}(t)$ and the particle number operators $\hat{N}_{\pm}(t)$
given by
\begin{eqnarray}
\Omega_{\pm}(t)&=&\Omega(t)\mp S(t)\;,\label{ompm}\\
{\hat N}_{\pm}(t)&=&{\hat a}^{\dag}_{\pm}(t){\hat
a}_{\pm}(t)\;.\label{nn}
\end{eqnarray}
Besides, one can observe that
the eigenvectors of Hamiltonian (\ref{hamquantosc}) can be written as
\begin{eqnarray}
|n_+,n_-,t> =
\frac{1}{\sqrt{n_+!}}\frac{1}{\sqrt{n_-!}}\left({\hat
a}^{\dag}_{+}(t)\right)^{n_+} \left({\hat
a}^{\dag}_{-}(t)\right)^{n_-}|0>\;,\label{state}
\end{eqnarray}
while the corresponding eigenvalues take the form
\begin{equation}
E_{n_+,n_-}(t) = \Omega_{+}(t) \left(n_+ + \frac{1}{2}\right) +
\Omega_{-}(t) \left(n_- + \frac{1}{2}\right)\;. \label{eigenvalues}
\end{equation}

Let us now consider an interesting situation such that
\begin{eqnarray}
\Omega(t) = S(t) = m\omega^2f_{\kappa_a}(t) +\frac{1}{f_{\kappa_a}(t)m}\;.\label{equality}
\end{eqnarray}
One can check that it appears  when functions $f_{\kappa_a}(t)$ and $g_{\kappa_a}(t)$ satisfy the following condition
\begin{eqnarray}
f_{\kappa_a}(t)\cdot g_{\kappa_a}(t) = 1\;.\label{condition}
\end{eqnarray}
Then, we have
\begin{eqnarray}
\Omega_{-}(t)=2\Omega(t)\;\;\;,\;\;\;\Omega_{+}(t)= 0
\;,\label{mamy}
\end{eqnarray}
and, consequently, the spectrum (\ref{eigenvalues}) provides the Landau energy levels with the time-dependent frequency
$\Omega(t)$\footnote{For canonical deformation $(f_{\kappa_1}(t) = \kappa_1)$ we get constant frequency $\Omega(\kappa_1)$.}
\begin{equation}
E_{n}(t) = \Omega(t) \left(2n + 1\right)\;\;;\;\;n = 0,1,2,3, \ldots \;. \label{landau}
\end{equation}

It should be mentioned, however, that the above result can be obtained in simpler (but less general) way as well. Firstly, one can observe
that under the condition (\ref{condition}) the  Hamiltonian (\ref{2dh1}) takes a particular form
\begin{eqnarray}
\hat{H} = \hat{H}(t) =
\frac{1}{2M(t)}\left({{{p}}_1^2}+{{{p}}_2^2} \right)  +
\frac{1}{2}M(t)\Omega^2(t)\left({{{x}}_1^2}+{{{x}}_2^2} \right)
- \Omega(t)L\;. \label{nowyham}
\end{eqnarray}
Next, if one defines the following creation and annihilation operators\footnote{$[\;\hat{A}(t),\hat{A}^{\dag}(t)\;] = 1$.}
\begin{eqnarray}
\hat{A}(t) = \frac{1}{\sqrt{2}}\left(\pi_1 + i\pi_2\right)\;\;\;,\;\;\;
\hat{A}(t) = \frac{1}{\sqrt{2}}\left(\pi_1 - i\pi_2\right) \;,\label{oper1}
\end{eqnarray}
with
\begin{eqnarray}
\pi_1(t) &=& \frac{1}{\sqrt{2M(t)\Omega(t)}}\left(p_1 + M(t)\Omega(t)x_2 \right)\;,\label{ped1}\\
\pi_2(t) &=& \frac{1}{\sqrt{2M(t)\Omega(t)}}\left(p_2 - M(t)\Omega(t)x_1 \right)\;,\label{ped2}
\end{eqnarray}
then, the function (\ref{nowyham}) can be written as
\begin{eqnarray}
\hat{H}(t) = \Omega(t)\left[\pi_1^2 + \pi_2^2\right] =
2\Omega(t)\left(\hat{A}^\dagger(t)\hat{A}(t) + \frac{1}{2} \right) \;. \label{nowyham1}
\end{eqnarray}
Consequently, it is easy to notice that (in fact)  the corresponding eigenvectors and the corresponding eigenvalues
are given by
\begin{eqnarray}
|n,t> &=&
\frac{1}{\sqrt{n!}}\left({\hat
A}^{\dag}(t)\right)^{n}|0>\;,\label{nowystate}\\
E_{n}(t) &=& \Omega(t) \left(2n + 1\right)\;\;;\;\;n = 0,1,2,3, \ldots \;, \label{nowylandau}
\end{eqnarray}
respectively.

\section{Final remarks}

In this paper we find the time-dependent energy levels for an oscillator model defined on the twisted N-enlarged Newton-Hooke space-time (\ref{spaces}).
Moreover, we  demonstrate that for a special choice of the deformation parameters of the corresponding phase space (see formula (\ref{condition})) the above spectrum can be identified with the time-dependent Landau one.

As it was already mentioned in Introduction the time-dependent Landau energies for quantum space (\ref{spaces}) (with function $g_{\kappa_a}(t)$ equal zero) has been already found in \cite{daszlandau}.
Precisely, there has been provided the energy levels for a nonrelativistic particle  moving in
uniform magnetic $(B)$ as well as in uniform electric (external) field; they read as follows
\begin{equation}
{\hat E}_{n}(t) = {\hat \Omega}(t) \left(2n + 1\right)\;\;;\;\;n = 0,1,2,3, \ldots \;, \label{daszlandau}
\end{equation}
with frequency ${\hat \Omega}(t)$ given by
\begin{equation}
 {\hat \Omega}(t) = \left(1 - \frac{f_{\kappa_a}(t)B}{4}\right)\frac{B}{2m}
\;, \label{daszfreq}
\end{equation}
where $m$ denotes the mass of particle.

Finally, it should be mentioned that the presented  investigation
has been performed for most general (constructed explicitly) type of space-time noncommutativity at nonrelativistic level.

\section*{Acknowledgments}
The author would like to thank J. Lukierski for valuable discussions.
 This paper has been financially  supported  by Polish
NCN grant No 2011/01/B/ST2/03354.


\begin{thebibliography}{99}
\bibitem{snyder}H.S. Snyder, Phys. Rev. 72, 68 (1947)
\bibitem{2}S. Doplicher, K. Fredenhagen, J.E. Roberts, Phys. Lett. B 331, 39 (1994);
Comm. Math. Phys. 172, 187 (1995); hep-th/0303037
\bibitem{2a}A. Kempf and G. Mangano, Phys. Rev. D 55, 7909 (1997);
hep-th/9612084
\bibitem{recent}A. Connes, M.R. Douglas, A. Schwarz, JHEP 9802, 003 (1998)
\bibitem{string1}N. Seiberg and E. Witten, JHEP 9909, 032 (1999);
hep-th/9908142
\bibitem{mech}A. Deriglazov, JHEP 0303, 021 (2003); hep-th/0211105
\bibitem{ddddd}S. Ghosh, Phys. Lett. B 648, 262
(2007)
\bibitem{qm}M. Chaichian, M.M. Sheikh-Jabbari, A. Tureanu, Phys.
Rev. Lett. 86, 2716 (2001); hep-th/0010175
\bibitem{prefield}P. Kosinski, J. Lukierski, P. Maslanka, Phys. Rev.
D 62, 025004 (2000); hep-th/9902037
\bibitem{field}M. Chaichian, P. Pre\v{s}najder and  A. Tureanu,
Phys. Rev. Lett. 94, 151602 (2005); hep-th/0409096
\bibitem{fiorewess} G. Fiore, J. Wess, Phys. Rev. D
75, 105022 (2007); hep-th/0701078
\bibitem{qg1}V. Chari, A. Pressley, \textit{"A Guide to Quantum
Groups"}, Cambridge University Press, Cambridge, 1994
\bibitem{qg3} L.A. Takhtajan, \textit{"Introduction to Quantum
Groups"}; in Clausthal Proceedings, Quantum groups 3-28 (see High
Energy Physics Index 29 (1991) No. 12256)
\bibitem{class1}S. Zakrzewski, \textit{"Poisson Structures on the Poincare
group"}; q-alg/9602001
\bibitem{class2}
Y. Brihaye, E. Kowalczyk, P. Maslanka, \textit{"Poisson-Lie structure on Galilei
group"}; math/0006167
\bibitem{nnh}M. Daszkiewicz, Mod. Phys. Lett. A27 (2012) 1250083; arXiv: 1205.0319 [hep-th]
\bibitem{oeckl}R. Oeckl, J. Math. Phys. 40, 3588 (1999)
\bibitem{chi}M. Chaichian, P.P. Kulish, K. Nashijima, A. Tureanu, Phys. Lett. B
604, 98 (2004); hep-th/0408069
\bibitem{dasz1}M. Daszkiewicz,
Mod. Phys. Lett. A 23, 505 (2008); arXiv: 0801.1206 [hep-th]
\bibitem{kappaP}J. Lukierski, A. Nowicki, H. Ruegg and V.N. Tolstoy, Phys. Lett.
B 264, 331 (1991)
\bibitem{kappaG}S. Giller, P. Kosinski, M. Majewski, P. Maslanka
and J. Kunz, Phys. Lett. B 286, 57 (1992)
\bibitem{lie1}
J. Lukierski and M. Woronowicz, Phys. Lett. B 633, 116 (2006); hep-th/0508083
\bibitem{qdef}O. Ogievetsky, W.B.  Schmidke, J. Wess, B. Zumino, Comm. Math. Phys.
150, 495 (1992)
\bibitem{paolo}
P. Aschieri, L. Castellani, A.M. Scarfone, Eur. Phys. J. C 7, 159
(1999); q-alg/9709032
\bibitem{romero}J.M. Romero and J.D. Vergara, Mod. Phys. Lett. A 18,
1673 (2003)
\bibitem{romero1}
J.M. Romero, J.A. Santiago, J.D. Vergara, Phys. Lett. A 310, 9
(2003)
\bibitem{cytowania}Y. Miao, X. Wang, S. Yu; arXiv: 0911.5227 [math-ph]
\bibitem{kijanka}A. Kijanka, P. Kosinski, Phys. Rev. D 70, 12702
(2004)
\bibitem{daszwal}M. Daszkiewicz, C.J. Walczyk, Phys. Rev. D 77, 105008 (2008)
\bibitem{toporzelek} E. Harikumar, A.K. Kapoor; arXiv: 1003.4603 [hep-th]
\bibitem{giri}P.R. Giri, P. Roy; 0803.4090 [hep-th]
\bibitem{pioneer}J.D. Anderson, P.A. Laing, E.L. Lau, A.S. Lin, M.M.
Nieto, S.G. Turyshev, Phys. Rev. Lett. 81, 2858 (1998)
\bibitem{hallcan}O.F. Dayi, A. Jellal, J. Math. Phys.  {43}, 4592 (2002);  hep-th/0111267
\bibitem{hallcan1}S. Dulat, K. Li, Chin. Phys. C 32, 92 (2008); arXiv: 0802.1118 [math-ph]
\bibitem{hallcan2}B. Basu, S. Ghosh, Phys. Lett. A 346, 133 (2005); cond-mat/0503266
\bibitem{hallcanf}A. Kokado, T. Okamura, T. Saito, \textit{"Noncommutative Hall Effect"}; hep-th/0210194
\bibitem{giri}P.R. Giri, P. Roy, Eur. Phys. J. C 57, 835 (2008); arXiv: 0803.4090 [hep-th]
\bibitem{daszlandau}M. Daszkiewicz,  Acta Phys. Polon. B 44, 59 (2013); arXiv: 1302.0827 [hep-th]
\end{thebibliography}
\end{document}